# Electrical Control of Broadband Terahertz Wave Transmission with Two-Terminal Graphene Oxide Devices

*Seungwoo Lee[1†*], Kyung Eun Lee[2†], Won Jun Lee[2], Byung Cheol Park[3], Byungsoo Kang[3], Euyheon Hwang[4*], and Sang Ouk Kim[2*]*

[1]SKKU Advanced Institute of Nanotechnology (SAINT) & School of Chemical Engineering, Sungkyunkwan University (SKKU), Suwon 16419, Republic of Korea

[2]National Creative Research Initiative Center for Multi-Dimensional Directed Nanoscale Assembly, Department of Materials Science and Engineering, Korea Advanced Institute of Science and Technology (KAIST), Daejeon 34141, Republic of Korea

[3]Department of Mechanical Engineering, Korea Advanced Institute of Science and Technology (KAIST), Daejeon 34141, Republic of Korea

[4]SKKU Advanced Institute of Nanotechnology (SAINT) & Department of Physics, Sungkyunkwan University (SKKU), Suwon 16419, Republic of Korea

**ABSTRACT**

Carbon nanomaterials such as carbon nanotubes and graphene have proved to be efficient building blocks for active optoelectronic devices. Especially, the exotic properties of crystalline graphene, such as a linear/gapless energy dispersion, offer a generic route to the development of active photonic modulator at the infrared (IR) and terahertz (THz) regime with large modulation depth. Here, we show that graphene oxide (GO), an oxygenated derivative of graphene with randomly distributed molecular defects (e.g., adsorbed water molecules and punched holes), can provide a different way to effectively control broadband THz transmission amplitude, when incorporated into two-terminal electrode devices. Electrically trapped charge carriers within localized impurity states (LIS) of GO, which originate from fully randomized defective structure of GO, results in a large modulation of transmission amplitude (~30%) for broadband THz waves (0.3 ~ 2.0 THz) even at room temperature. Interesting hysteretic behavior observed in the control of broadband THz transmission further confirms the key role of trapped charge carriers in switching of broadband THz waves. The device architecture constructed by simple solution printing of GO onto the two-terminal electrode enables easy-to-implement active photonic devices.

## INTRODUCTION

Over the last decade, major technological innovations related to graphene, single crystalline layer of hexagonally arranged carbon atoms, have transformed various research fields such as electronics, energy, and photonics[1-11]. More specifically, emerging graphene optoelectronic technologies have exploited a possibility of electrical control on the interaction between electromagnetic waves and graphene[9,12-18]. Indeed, several unique features including (i) linear and gapless energy dispersion and (ii) electrically adjustable carrier density[19,20] allow graphene to be of central importance to a rapidly expanding frontier in the construction of deep-subwavelength-scaled, active photonic device for the fast and efficient modulation of broadband electromagnetic waves (e.g., from mid-infrared (IR) to terahertz (THz))[7].

These recent advances raise a fundamental question on how defective GO, where 1-2 nm size graphene islands are surrounded by defective amorphous oxygenated domains with hydroxyl/ketone/epoxide functional groups,[21] should be different from highly crystalline graphene particularly in terms of active control of light-matter interaction. Although a few experimental works on THz spectroscopy of GO have been recently reported[22], there is still neither report on the active control of THz waves using GO nor the studies on the effect of GO complex defects on THz spectroscopy.

In this work, we aim to empirically exploit the role of GO's complex molecular defects in electrical switch of broadband THz waves at room temperature. Notably, we found that the amount of the trapped charge carriers within the fully randomized, complex molecular defects (i.e., localized impurity states (LIS)) of GO can be precisely adjusted by the dc biasing with versatile and simple device architecture (i.e., a two-terminal, interdigitated electrode device); the excitation of the electrically trapped charge carrier in turn leads to a large change in broadband THz transmission (e.g., ~ 30 % at 1.0 THz). This dramatic change in broadband THz transmission (i.e., 0.3 ~ 2 THz) was achieved with a deep-subwavelength (< λ/10,000) thick GO paper (~ 500 nm) without any assistance of the foreign structural motifs (e.g., metamaterials, plasmonic structures, and Fabry-Pérot architectures). Systematic THz spectroscopy of the chemically reduced GO further revealed the key role of GO's complex molecular defects in the active control of broadband THz transmission. Also, we observed the trapped charge carriers within LIS of GO could additionally provide interesting hysteretic behavior in the control of broadband THz transmission, which could be applied to the THz based photo-memory devices.

## RESULTS & DISCUSSION

Figures 1a-c present schematic and macro/microscopic images of the two-terminal GO devices used in the present work. The GO device mainly consists of two functional parts: (i) GO paper (i.e., multi-stacked GO flakes) and (ii) two-terminal, interdigitated gold (Au) electrodes (Au microwire array) patterned on the polyimide (PI) flexible substrate (2.6 μm in thickness). High quality monolayer GO flakes, prepared by acidic chemical exfoliation of graphite[23], were dispersed in deionized water; then, GO aqueous solutions were vacuum-filtrated in order to obtain mechanically robust, free-standing paper platform (~ 500 nm in thickness). The transmission electron microscope (TEM) image and UV/VIS spectroscopy of the obtained GO flakes are summarized in Supporting Information (Figure S1 and S2). In particular, UV/VIS spectroscopy and its Tauc plot analysis verify the presence of both energy bandgap (3.4−4.0 eV)

and LIS, which contrast to graphene with zero bandgap. The mechanical robustness of the paper-type film can facilitate the transfer of GO onto any desired substrate; herein, by using solvent-assisted transfer printing (see Supporting Information, Figure S3), GO paper was directly integrated with the interdigitated, two-terminal electrodes (100 nm thick Au micro wire array with 10 nm Ti adhesion layer), as shown in Figures 1b-d. Scanning electron microscope (SEM) image in Figure 1d shows the large-area GO paper (1 cm by 2 cm), conformably printed onto the electrodes. The chemical composition of the GO papers was characterized by X-ray photoemission spectroscopy (XPS) (see Figure 1e). Appearance of O-C-O/-OH (at 286.7 eV) and C=O (at 288.2 eV) shows successful oxidation of graphite and the atomic fractions of carbon and oxygen in GO paper, calculated by the intensity of each peak, were found to be 64.89 % and 35.11 %, respectively.

The electrode architecture on the PI substrate was developed by consecutive processes consisting of conventional photolithography, metal evaporation (electron beam deposition), and the lift-off processes. The width of the Au microwire electrodes was 5 μm and the gap between the two-terminal electrodes was varied between 20 μm and 100 μm. This two-terminal electrode structure was designed for the transparency at broadband THz waves of interest (0.3 − 2.0 THz), which are linearly polarized along the direction perpendicular to the grating vector of Au microwire array (referred to as THz transparent electrode, TTE), as presented in Figure S4, Supporting Information). More significantly, the two-terminal "interdigitated" electrode was designed to effectively increase the electrode area over the large-area GO paper, compared with a simple two-terminal device (see the simulated electrical potential in the vicinity of the TTE in Supporting Information, Figure S5). THz time-domain spectroscopy (THz-TDS) was used to characterize the THz transmission through the two-terminal GO device[17]. All of the THz-TDS measurements were performed at room temperature (15 ~ 20 °C), and the polarization of the THz wave was set to be perpendicular to the Au microwire array electrodes for the transparency at the THz frequency of interest (0.3 ~ 2.0 THz). Additionally, a custom-built encapsulation box for the THz-TDS setup with nitrogen gas purging provided the ability to measure the THz transmission under low relative humidity (as low as 1 %), which allowed us to minimize the moisture-assisted, permanent electrochemical reduction of GO[24,25].

The first subject in the present work involved assessing whether an electrical dc biasing of GO can control THz wave transmission amplitude. As described previously, the systematic THz spectroscopy analysis of GO with various levels of dc biasing was aided by the interdigitated two-terminal electrodes (20 μm gap of interdigitated electrodes). The dc bias voltage ($V_{ds}$) was quasi-statically swept from − 0.1 V to + 0.1 V: for $| V_{ds} | > 0.1$, the permanent electrical reduction of GO occurred[24,25]. The dwell time per each voltage was 250 s. Figures 2a (a negative dc biasing) and 2b (a positive dc biasing) show the representative control of THz transmission by dc biasing of GO paper. Two important features are noteworthy. First, the THz transmission of the as-prepared GO (i.e., neither reduced nor doped GO) decreases monotonically with frequency. This behavior is completely attributed to the presence of the GO. The dashed lines in Figure 2 show the THz transmission for the substrate only (i.e., without GO). As shown in Figure 2, the most of the THz wave is transparent to the substrate. Thus, the THz waves are absorbed into the GO paper and this absorption increases with frequency. This behavior (monotonic decrease of transmission amplitude with THz frequency) is different from that for the reduced GO. It is well known that in the highly reduced GO, the transmission of THz wave is generally independence of frequency[22]. Second, the THz transmission of GO is continuously decreased, as $|V_{ds}|$ is changed from 0 V to -0.1 V or + 0.1 V. In particular, the broadband THz transmission change by dc biasing (i.e., $\Delta T/T_{V=0}$, where $\Delta T = T_{V=0} - T$) was observed up to ~ 30 % (e.g., at 1.0 THz) in the frequency range between 0.3 and 2.0 THz (for the positive dc biasing).

We then quantified the behavior of $\Delta T/T_{V=0}$ at 1.0 THz, in more depth, as shown in Figure 3a. The two distinct regimes are clearly observed: (i) steep increase of $\Delta T/T_{V=0}$ for $|V_{ds}|<0.005V$ and (ii) slow increase of $\Delta T/T_{V=0}$ for $|V_{ds}|>0.005V$. We also measured channel current of the GO paper between two electrodes with respect to $V_{ds}$ (Figure 3b). Interestingly, we observed that current is almost zero during a steep increase of $\Delta T/T_{V=0}$, whereas the current shows ohmic behavior during a moderate increase of $\Delta T/T_{V=0}$ (for $|V_{ds}|>0.005V$). Meanwhile, the anisotropy in electrical conduction within GO (e.g., positive dc biasing > negative dc biasing) was often observed in our experimental device because of the non-uniformly distributed contact of electrodes and GO paper at the two terminals[26]. The non-identical behaviors of the change in $\Delta T/T_{v=0}$ between positive and negative dc biasing could be attributed to this anisotropic electrical conduction of our devices. Thus, it is natural to ask, 'why does dc biasing of GO resulted in a decrease in the broadband THz transmission?'

A speculative explanation derived from these observations is based on two things. (i) The dc biasing can provide electrons into the device; subsequently, the complex molecular defects of GO can trap these injected electrons within LIS (see Figure 3c). (ii) The electrons, trapped within randomly distributed LIS, can be excited to the extended region of the conduction band (or upper energy level) by broadband THz irradiation. In the present work, the bandgap energy of GO was measured to be approximately 3.4 − 4.0 eV (see Figure S2b). Thus, the bandgap of GO is nothing to do with the THz absorption. However, the long band tail in the UV/VIS Tauc plot of GO aqueous solution indicates the presence of randomly distributed LIS (at the bottom of the conduction band and at the top of valence band). In the trapping regime (Figure 3c), the transporting electrons driven by $V_{ds}$ should be first accumulated into the fully randomized LIS, in that the electrical conductivity between drain and source electrodes cannot be increased significantly (see Figure 3b). The hopping current through the impurity states may dominate for $|V_{ds}|< 0.005$ V. However, the trapped electrons within LIS can be excited into the extended states of the conduction band (i.e., above the mobility edge) through broadband THz absorption, as evidenced by a significant increase in $\Delta T/T_{V=0}$ despite of negligible change in current (see Figure 3a). As $V_{ds}$ increases (especially beyond $V_{ds}=0.005$), the trapped electrons within LIS become saturated gradually, while the electrons can be transportable across the GO paper (see Figure 3d of transporting regime). This in turn results in both the increase in conductance (Figure 3b) and a saturated enhancement of $\Delta T/T_{V=0}$ beyond $V_{ds}$ of 0.005.

Specifically, compared with single layer graphene, the GO paper used in this study can be treated as a relatively thick, amorphous carbon layer (at least 1000 layers of the stacked GO flakes), in which quite complex and various molecular defects such as hydroxyl/ketone/epoxide functional groups, vacancies, and the adsorbed water are fully randomized. This unique randomness of the molecular defects makes the LIS of GO paper almost continuous and broadband. Thus, in a sense, broadband THz waves can be absorbed through the excitation of the trapped electrons within nearly continuous and broadband LIS. Furthermore, the amount of the trapped electrons within the LIS of GO paper can be tuned by dc biasing, so as to achieve active control of broadband THz transmission.

We next characterized the hysteretic behavior of the THz transmission change by cyclic dc biasing of GO paper, as summarized in Figure 4. As such, the relationship between trapped electrons and THz transmissions can be systematically verified in more depth. To demonstrate the hysteretic behavior of the THz transmission change by dc biasing of GO, we measured the transmission of the two-terminal GO device at 1.0 THz during cyclic $V_{ds}$ sweep. The swept voltage sequence was as follows: 1st sweep from 0 V to – 0.1 V, 2nd sweep from – 0.1 V to 0 V, 3rd sweep from 0 V to + 0.1 V, and 4th sweep from + 0.1 V to 0 V. During each sweep, the THz transmission of GO was measured at fixed voltages. The dwell time per each voltage was 250 s. The measurement of THz transmission versus the voltage forms a hysteresis loop. The obtained hysteresis loop of THz wave transmission at 1.0 THz and the relevant current change are summarized respectively in Figure 4a and 4b; Figure 5 summarizes the proposed underlying mechanism for the hysteresis loop of $\Delta T$.

When the voltage decreases from 0 V to -0.1 V, the most of electrons from the electrode are trapped in the available impurity states, which gives rise to the reduction of THz transmission, because the trapped electrons absorb THz wave. For $V_{ds}$ < -0.005V all electrons from the electrode accumulate in the GO channel (Figure 5a-b), as evidenced both by the dramatic decrease in THz transmission (Figure 4a) and by negligible current (Figure 4b). Then, as this accumulation is saturated gradually, charges become transportable across the GO paper (Figure 5c), resulting in the increase in current. When the voltage increases from – 0.1 V to 0 V the THz transmission decreases further instead of going back to the initial point at $V_{ds}$=0 V (Figure 4a). This behavior can be understood from the trapped electrons, because the electrons cannot be released from the impurity sites during the negative dc biasing. Indeed, the THz transmission was found to be further reduced, during 2nd sweep (Figure 4a). In contrast, current showed the tendency of having back to the original point, owing to the gradually reduced dc bias (Figure 4b). The relatively weak hysteresis of current was well revealed as a direct evidence of electron trapping within LIS. After 2nd sweep (returning to 0 V), much more charges can be trapped within LIS of GO, compared with intrinsic GO (Figure 5d). As a result, THz transmission can be dramatically dropped (e.g., from 96 % to 64 % at 1.0 THz; 33 % of $\Delta T/T_{initial\ state}$, where $\Delta T$ is $T_{initial\ state} - T$) (Figure 4a).

In the reverse direction of the voltage (i.e., from 0 V to + 0.1 V and from + 0.1 V to 0 V), the potential of electrode is flipped with respect to 1st and 2nd sweeps, and the trapped electrons can be released. Especially at 0.1 V of 3rd sweep, this releasing process of trapped electrons (Figure 5e) can be facilitated, as evidenced by the increase in THz transmission again (see Figure 4a). Eventually, most of the trapped electrons was released during 4th sweep (Figure 5f), in that THz transmission was found to be returned almost to the original point after 4th sweeps (Figure 4a). Conclusively, the complex hysteretic loop of $\Delta T$ was achieved. As well as electron trapping, the quite complex chemical and structural defects of GO possibly lead to the complicated polarization behavior, contributing to such unique hysteric loop of $\Delta T$. This aspect needs to be more carefully investigated in the future works. Interestingly, such hysteretic characteristic allows us to further reduce the THz transmission amplitude, so as to further increase the available range of $\Delta T/T_{V=0}$ (e.g., for negative dc biasing, from 18 % to ~ 30 % at 1.0 THz).

This hysteretic behavior and the range of $\Delta T/T_{V=0}$ during cyclic $V_{ds}$ sweep should be dependent on the effective electrical potential. Indeed, the lowered electrical potential across the two-terminal electrodes (i.e., 100 μm gap of interdigitated electrodes) resulted in the reduced range of $\Delta T/T_{V=0}$ (during 1st and 2nd sweeps, ~ 15 %), as shown in Figure 6. This result indicates that the electron trapping and the relevant electrical control of THz transmission can be controlled by adjusting the effective electric potential between the interdigitated electrodes.

To further test the role of complex molecular defects in the electrical control of broadband THz transmission, we finally performed the THz spectroscopy of hydrazine ($NH_2NH_2$)-reduced GO (HReGO)[33,34]. As previously reported, the oxygen molecular defects with various bonding motifs (e.g., hydroxyl, carboxyl, and ketone groups) were effectively removed from solid-state GO (see details of HReGO, presented in Figure S6-S7, Supporting Information)[33]. Thus, this HReGO paper, in which the number of layers is almost the same with that for the as-prepared GO paper, can be treated as a good reference to directly investigate the effect of molecular defects on the trapped electron-mediated control of THz transmission. Herein, we systematically adjusted the hydrazine treatment time (20 sec, 60 sec and 180 sec). Figure 7a presents typical THz spectroscopic results of GO and HReGO with a controlled chemical reduction. The hydrazine treatment of GO resulted in the broadband reduction of THz transmission together with the distinctly flatten profile, since making GO significantly reduced[22]. Also, the degree of the reduction in THz transmission was proportional to the hydrazine treatment time (free carriers can be excited by intraband transitions).

More importantly, in the case of hydrazine-treated GO, the available range of $\Delta T/T_{V=0}$ during cyclic dc biasing was significantly narrowed (at 1.0 THz), as presented in Figure 7b. As with the results of Figure 4, we carried out cyclic $\Delta T/T_{V=0}$, as following sequence: 1st sweep from 0 V to – 0.1 V, 2nd sweep from – 0.1 V to 0 V, 3rd sweep from 0 V to + 0.1 V, and 4th sweep from + 0.1 V to 0 V. For this study, we used 20 sec hydrazine treated GO, as other hydrazine-treated GO papers (the hydrazine treatment times were 60 sec and 180 sec) were not dc biased due to the near-metallic properties (i.e., Figure S8, Supporting Information). Such significantly narrowed range of $\Delta T/T_{V=0}$ after the hydrazine treatment originated from the removed molecular defects and thus less efficient electron trapping; further elucidating the role of GO's molecular defect in the electrical control of THz transmission.

**CONCLUSTION**

The concept reported in the present work newly establishes a baseline for the pivotal role of GO's complex molecular defects in dc biasing-controlled THz transmission. The THz spectroscopy of dc biased GO together with cyclic I-V measurements were systematically performed to support the possibilities for the electrical control of broadband THz transmission. Therefore, the capabilities for switching THz transmission are more accessible with the simple and efficient device architecture (the conformably printed GO paper onto two-terminal, interdigitated electrode). This generic strategy should expand the range of possible active optoelectronic devices with a broadband and large change of THz transmission. Further improvements in the THz transmission change together with detailed investigations of switching time which will be critical for enabling practical application are possible; the potential steps in this direction include the optimization of the GO thickness, chemical composition, and device architectures.

**ACKNOWLEDGMENT**

We thank B. Min, J.-H. Ko, Y.-H. Kim for helpful discussion. This work was supported by Basic Science Research Program through the National Research Foundation of Korea (NRF-2014R1A1A2057763, funded by the Ministry of Education, Science, and Technology, Korea), the Pioneering Nano-Based Convergence HRD Center (BK21+ program at Sungkyunkwan University), the Basic Science Research Program (2009-0083540) of the National Research Foundation of Korea (the 8th National Core Research Center at Sungkyunkwan University, named by Center for Human Interface Nano Technology, HINT), and the Multi-Dimensional Directed Nanoscale Assembly Creative Research Initiative (CRI) Center (2015R1A3A2033061) of the National Research Foundation of Korea (MSIP).

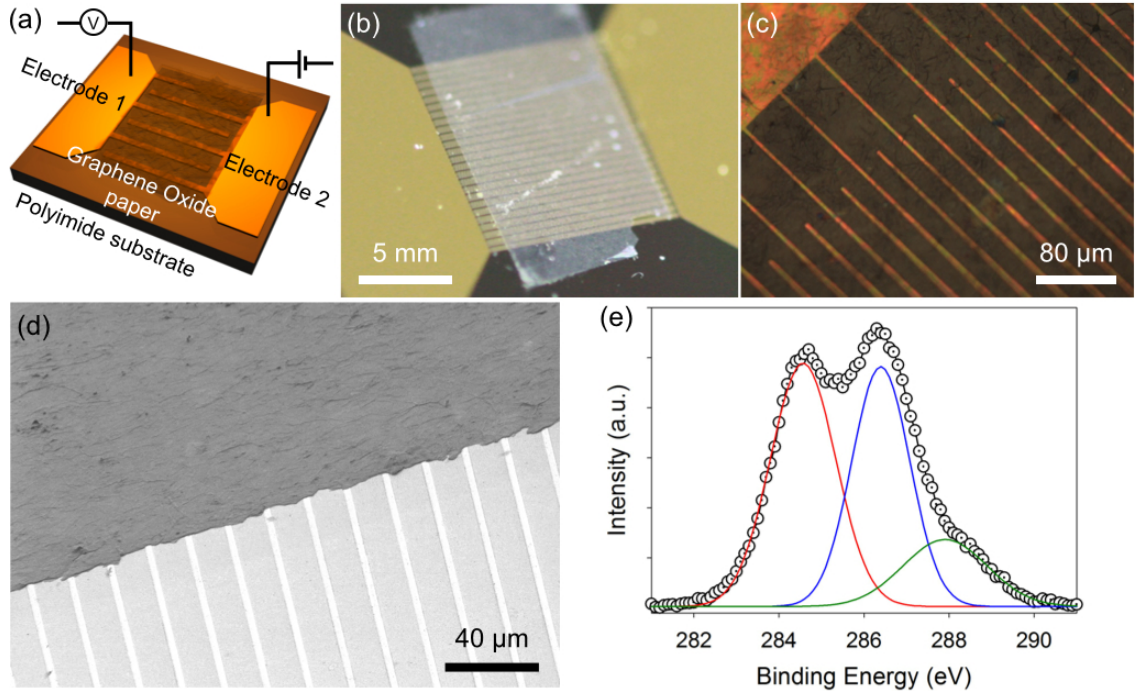

**Figure 1.** Two-terminal graphene oxide (GO) device for the electrical control of broadband terahertz (THz) transmission. (a) Schematic illustration of the two-terminal GO device architecture composed of an interdigitated electrode and GO paper (2 cm x 1 cm area and 500-nm thickness). The two-terminal electrodes (100 nm gold and 10 nm Cr) with an interdigitated geometry (gap of 20 μm and 100 μm) were fabricated onto polyimide (PI) film (2.7 μm thickness) by photolithography and the metal lift-off process. Then, the GO paper was printed onto the electrode. The electrode was designed to have an interdigitated geometry in order to increase the uniformity of the electrical bias over the large area of the GO paper. (b) Photograph of the integrated GO device. We can clearly see the whitish color of the pristine GO paper printed onto the two-terminal, interdigitated electrodes. This whitish color is the evidence of the non-reduced, pristine GO. (c) Optical microscope (OM) image (reflective mode) of one end of the GO paper/two-terminal, interdigitated electrodes (20 μm gap). (d) Top view scanning electron microscope (SEM) image of GO paper/two-terminal, interdigitated electrodes (20 μm gap). The GO paper was conformably printed onto the interdigitated electrodes. (e) The measured and fitted X-ray photoelectron spectroscopy (XPS) results of pristine GO. The XPS fitting was performed by Gaussian functioning of three main parts: C=C ($sp^2$ carbon at 284.6 eV), O-C-O/-OH (at 286.7 eV), and C=O (at 288.2 eV). The representative oxygen groups including the ketone, carboxyl, hydroxyl, and epoxide groups are well resolved.

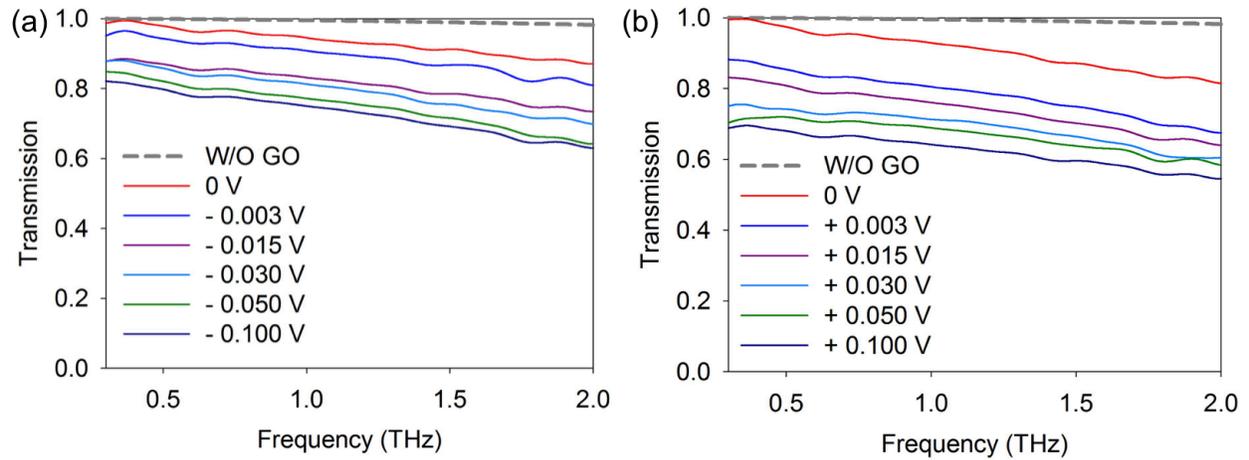

**Figure 2.** Representative THz spectroscopy of as-prepared GO paper (original GO paper) and switching of broadband THz transmission by dc biasing of GO. Each panel shows the THz transmission as a function of frequency for various negative (a) and positive (b) dc bias voltage. The dashed lines indicate the THz transmission of the substrate. The transmission of the as-prepared GO paper decreases with frequency. The dc biasing of GO enabled the active change in broadband THz transmission (0.3 ~ 2.0 THz).

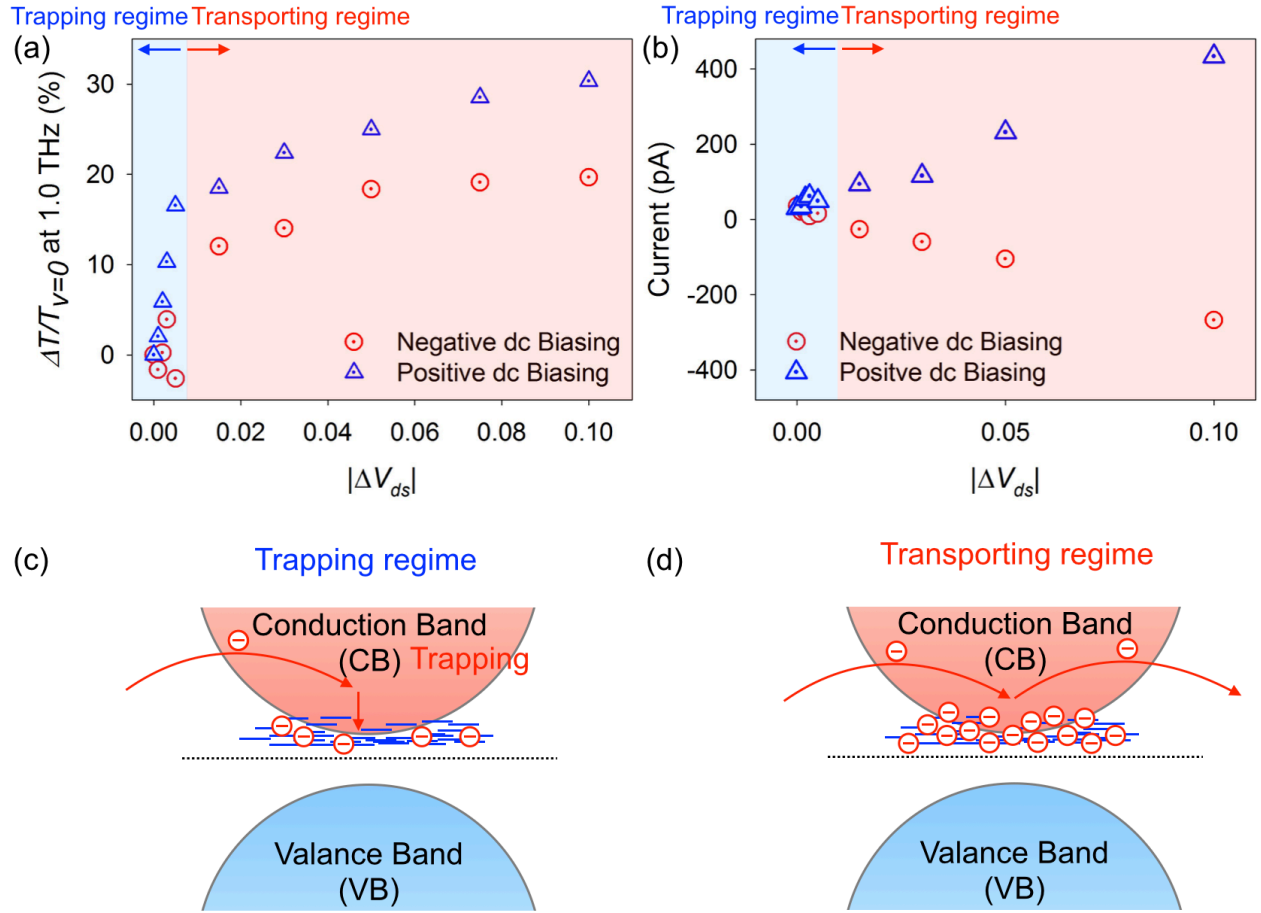

**Figure 3.** (a) THz transmission change ($\Delta T/T_{V=0}$ at 1.0 THz, where $\Delta T$ is $T_{V=0}$ -$T$) of two-terminal GO device as a function of $|V_{ds}|$ (the gap between interdigitated electrodes was 20 μm). Two distinct regimes are clearly observed including (i) a radical increase of $\Delta T/T_{V=0}$ under 0.005 of $V_{ds}$ (referred to as trapping regime) and a slow increase in $\Delta T/T_{V=0}$ beyond 0.005 of $V_{ds}$ (referred to as transporting regime). (b) Measured electrical current of GO as a function of $|V_{ds}|$; as with $\Delta T/T_{V=0}$, trapping and transporting regimes are obvious. (c) Schematic trapped electrons at low bias, in which the electrons driven by dc biasing are preferably trapped within localized impurity states (LIS) at the bottom of the conduction band (i.e., below the mobility edge). (d) Schematic for transporting regime at high bias, in which the electrons can be transported across GO beyond the saturation of LIS with electrons. Note that the impurity states at the top of the valence band are fully occupied and the majority current comes from electrons.

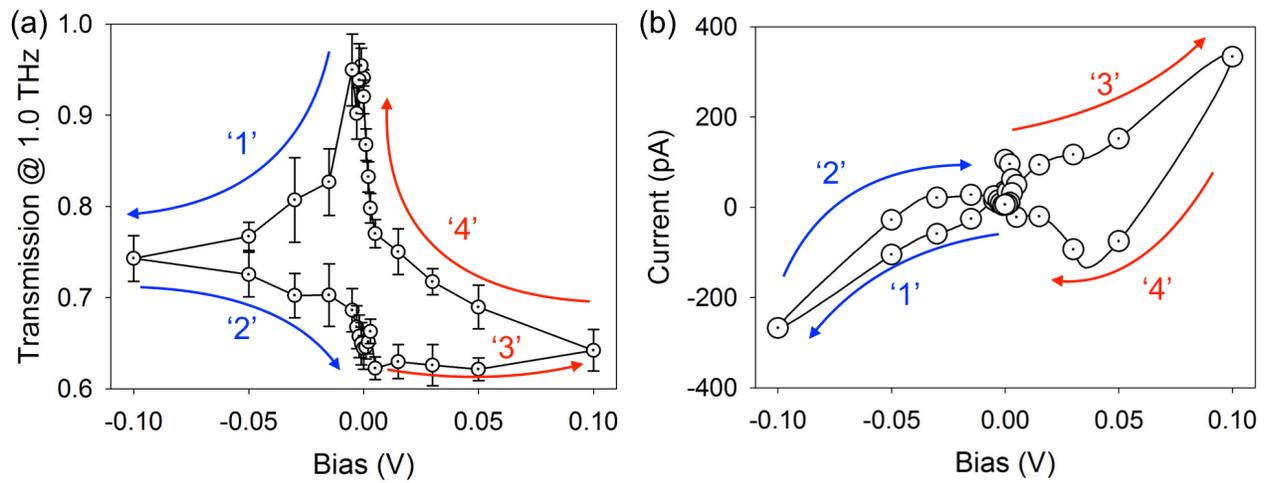

**Figure 4.** (a) Hysteretic behavior of THz transmission (1.0 THz) of GO during a cyclic change of dc bias voltage. The gap of interdigitated electrodes was 20 μm. (b) Representative cyclic I-V curve of GO. The hysteric behavior of I-V curve is a direct evidence of the presence of LIS and the relevant electron trapping within LIS.

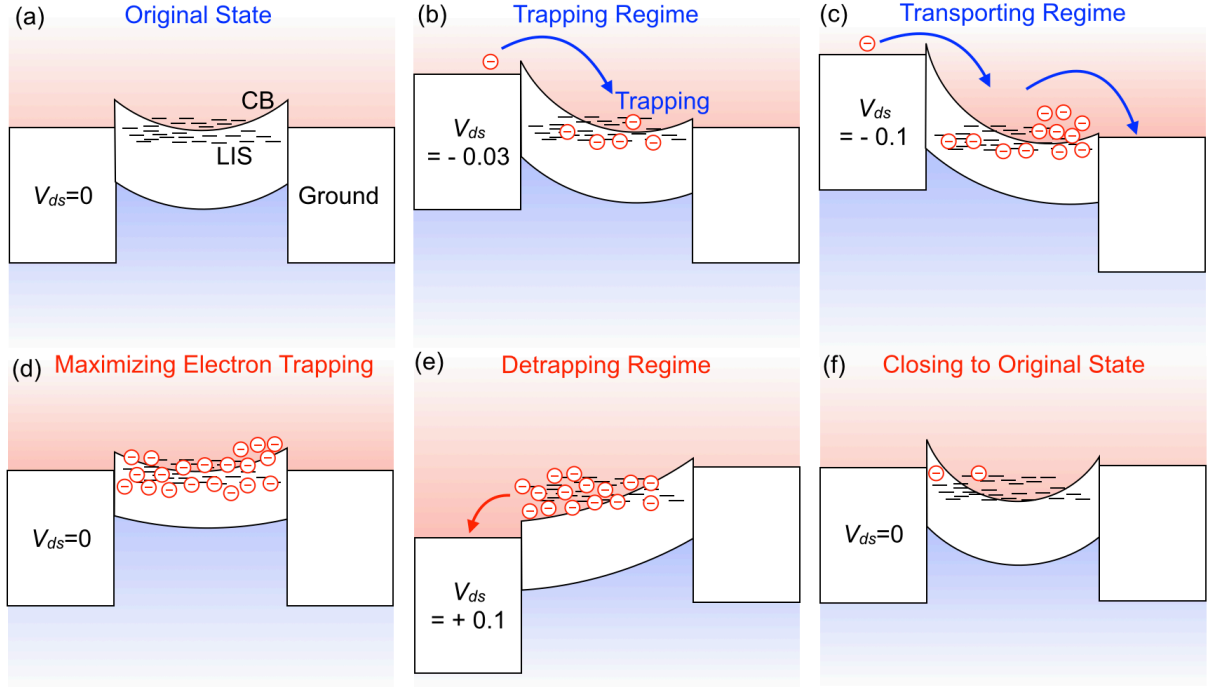

**Figure 5.** (a-f) Band diagrams of two-terminal GO devices during cyclic dc biasing (1st sweep from 0 V to – 0.1 V, 2nd sweep from – 0.1 V to 0 V, 3rd sweep from 0 V to + 0.1 V, and 4th sweep from + 0.1 V to 0 V.). A series of band diagrams proposes the underlying mechanisms for the hysteric change in THz transmission (1.0 THz) of GO during a cyclic change of dc bias voltage (the dwell time for each dc bias voltage was 250 s). (a) The initial state of GO with the presence of LIS, which is implemented into two-terminal device. (b) During 1st sweep, charges are first trapped within LIS below $V_{ds}$ of 0.005; (c) then, electrons can start to be transported across GO together with further electron trapping in LIS. During 2nd sweep, the negative dc bias is still effective in GO; thus, the electron trapping can be preceded further. (d) After 2nd sweep, the amount of the trapped electrons within LIS becomes maximized and THz transmission gets minimized. (e) During 3rd sweep, the energy band gets flipped, compared with 1st and 2nd sweeps; consequently, the trapped electrons can be released. This detrapping process can be facilitated at + 0.1 $V_{ds}$ of 3rd sweep, as evidenced by the increase in THz transmission again (see Figure 4a). (f) During 4th sweep, the detrapping of electrons is induced further; in the end, most of LIS becomes closed to empty state as with the original state.

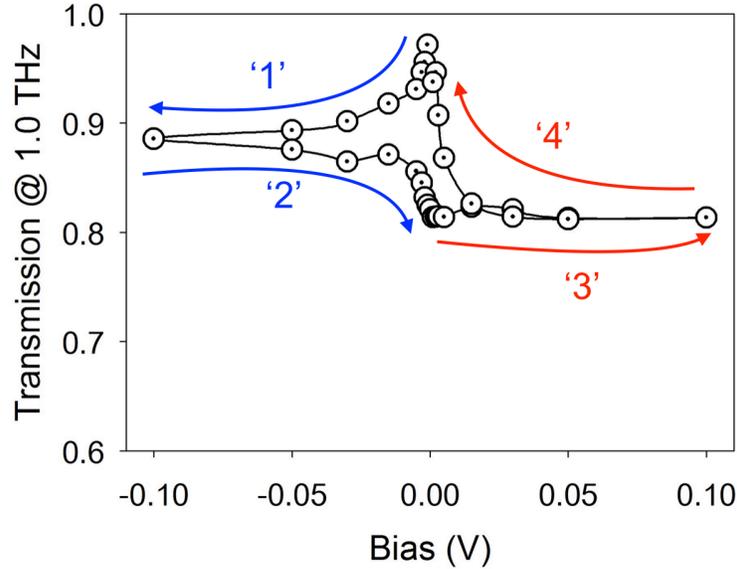

**Figure 6.** Hysteretic behavior of GO THz transmission (1.0 THz) during a cyclic dc biasing with 100 μm gap between interdigitated electrodes. Due to the much larger gap between interdigitated electrodes, the effective electrical potential was significantly reduced, compared with the results of Figure 4.

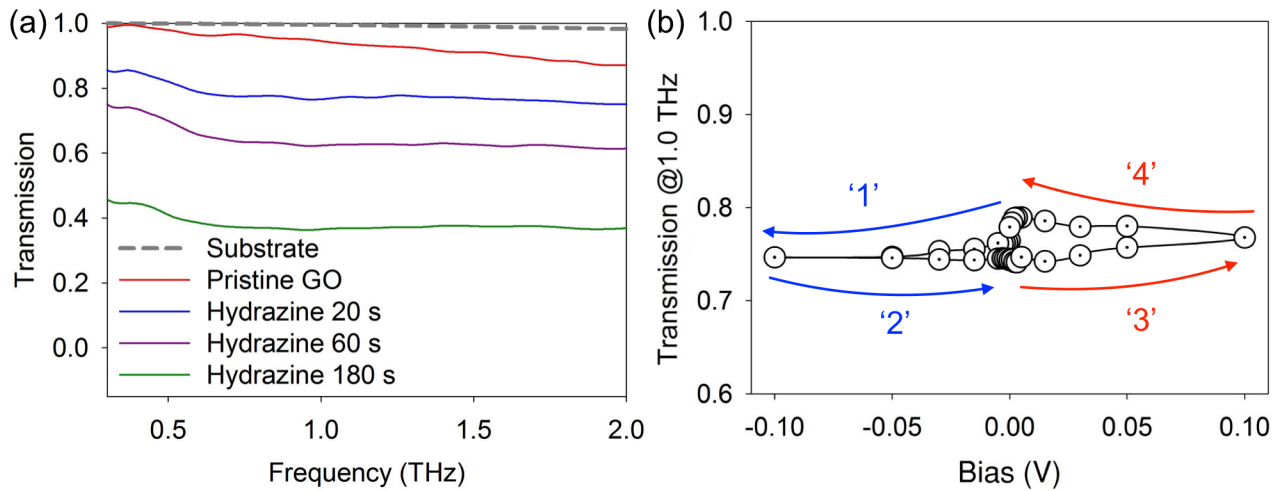

**Figure 7.** The effects of molecular defects (i.e., oxygen molecular derivatives) on the electrical control of THz transmission of GO. (a) THz spectroscopy of GO with controlled hydrazine treatment times. As the oxygen-related molecular defects are removed from GO by hydrazine treatment, the broadband THz transmission was significantly reduced, as with previous results[22]. (b) Cyclic measuring THz transmission of hydrazine-treated GO (for 20 sec) during cyclic dc biasing. The gap between interdigitated electrodes was 20 μm.

**Supporting Information for**

**Electrical Control of Broadband Terahertz Wave Transmission with Two-Terminal Graphene Oxide Devices**

**1. Transmission Electron Microscope (TEM) of Graphene Oxide (GO)**

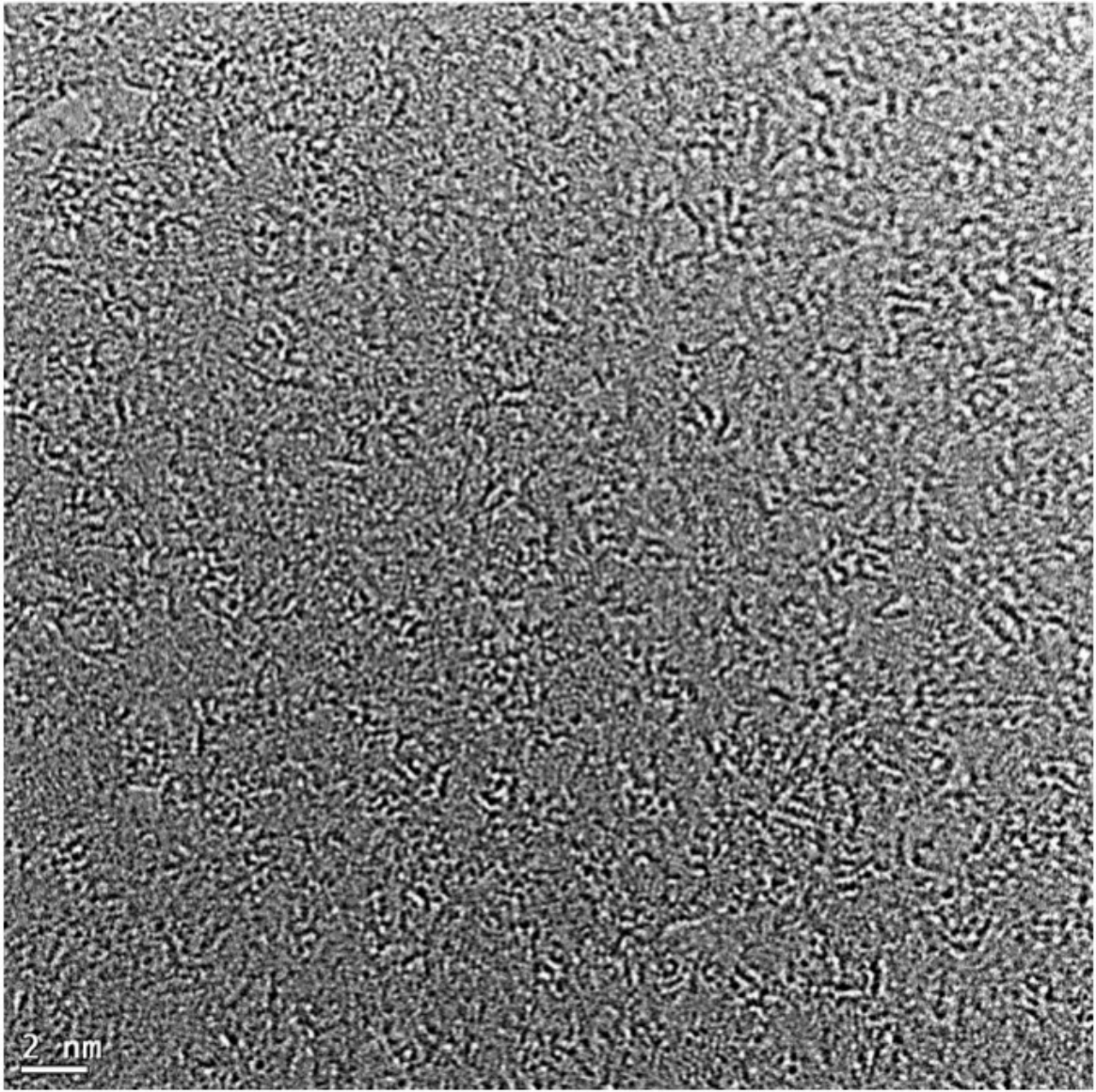

**Figure S1.** Transmission electron microscope (TEM) image of as-synthesized graphene oxide (GO) used in the present study.

## 2. UV/VIS spectroscopy of GO

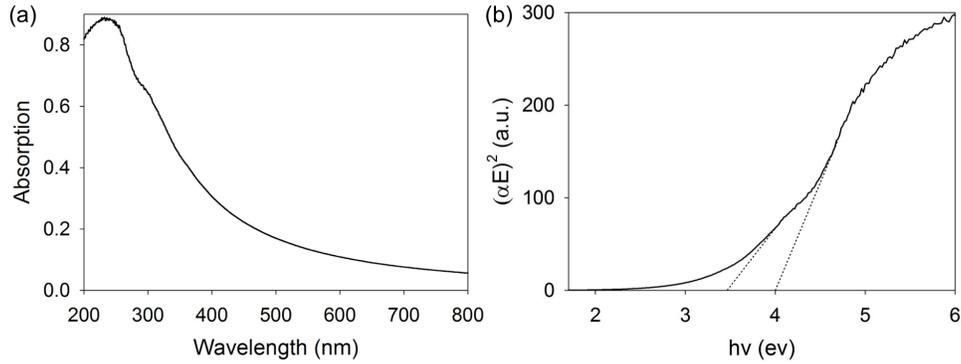

**Figure S2.** (a) UV/VIS spectroscopy and (b) Tauc plot of GO dispersed in deionized (DI) water. The absorption at *n-pi\** shoulder peak (320 nm wavelength) of UV/VIS indicates C=O functional groups of GO. Tauc plot (fitting line, highlighted by black dots) was used as for calculating band gap of GO (3.4 ~ 4.0 eV); the long band tail confirms the presence of localized impurity states (LIS).[S1]

## 3. Fabrication of two-terminal GO device

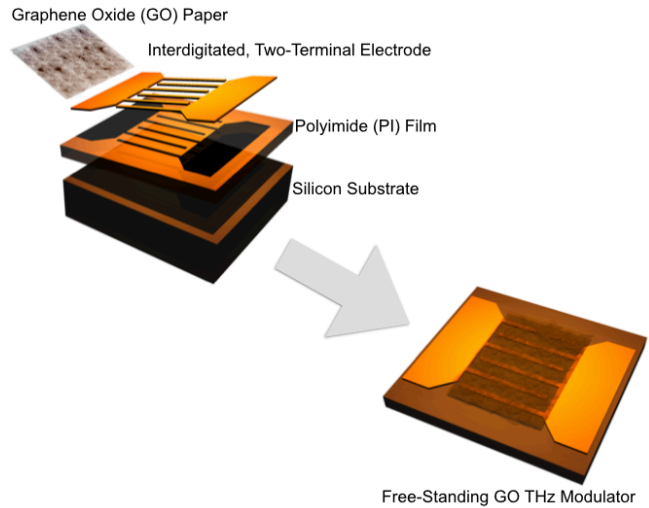

**Figure S3.** Schematic illustration for the fabrication of two-terminal GO device.

GO aqueous solution was vacuum-filtrated using polyvinylidene fluoride membrane filter (Millipore, 47 mm in diameter, 0.45 mm in pore sizes) and dried completely. After cutting into sizes of choice, the membrane was rapidly immersed into acetone-water mixture (7:3 ratio) to dissolve the membrane. Finally, floating GO paper was transferred on the two-terminal electrodes, fabricated onto polyimide (PI) film (see Figure S3). The film was rinsed with acetone to remove impurities.

**4. Transmission of two-terminal, interdigitated electrodes in the range from 0.3 to 2.0 THz**

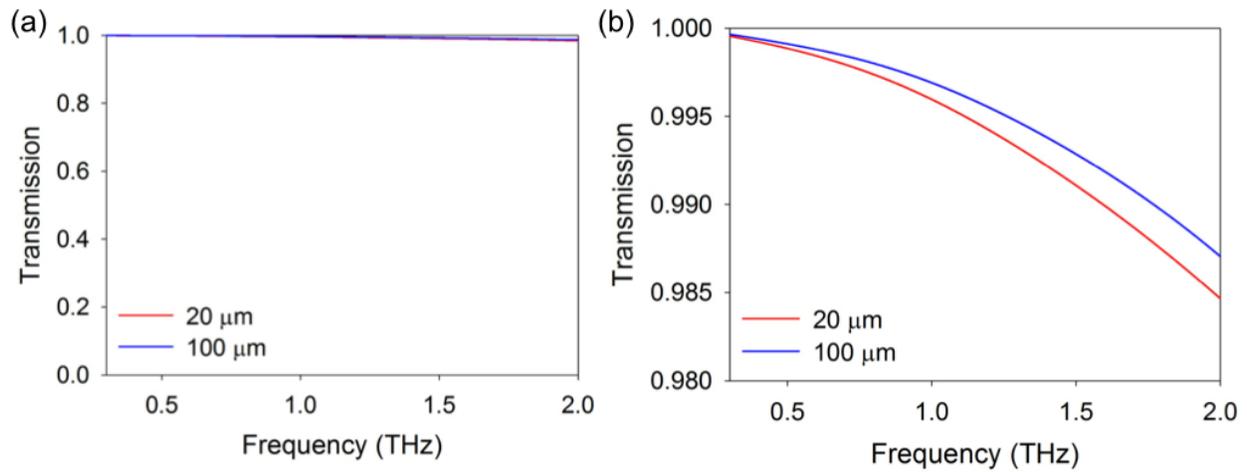

**Figure S4.** Simulated THz (0.3 ~ 2.0 THz) transmission of two-terminal, interdigitated electrode: (b) is magnified (a). The electric polarization of THz wave was set to be perpendicular to the line of two-terminal electrodes.

## 5. Simulated electric potential of interdigitated, two-terminal electrodes

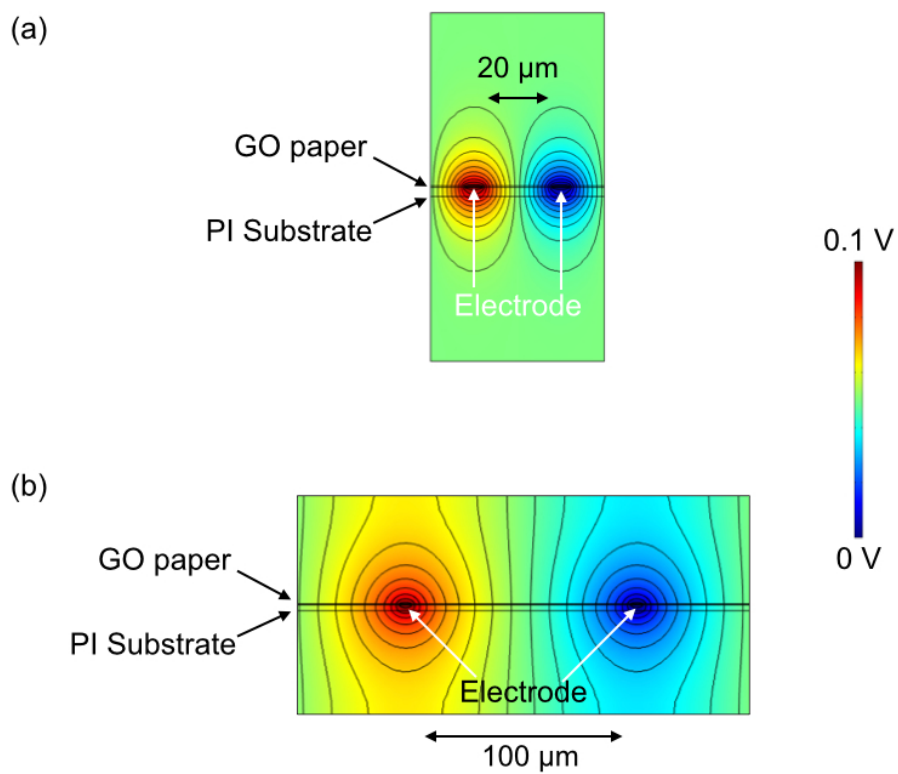

**Figure S5.** Simulated electric potential of 20 μm (a) and 100 μm (b) gap of interdigitated two-terminal electrode: right electrode in here is grounded.

## 6. Details of the chemically reduced GO

The chemical compositions of GO and HReGO are compared in the XPS results of Figure S6; the intensity and area under the peak corresponding to O-H and O-C-O bonds were significantly reduced with the hydrazine treatment time (in addition, the reduction of the shoulder peak of UV/VIS absorption at ~ 300 nm indicating *n-pi$^*$* transitions of C=O functional groups[S2] is presented in Figure S6). As previously reported, the nitrogen moiety from hydrazine preferably attaches to the edges of HReGO rather than on basal plane while the oxygen molecules with various bonding motifs (e.g., hydroxyl, carboxyl, and ketone groups) are effectively removed[S3]. More importantly, it is generally known that the hydrazine treatment allows GO to restore the graphitic *sp$^2$* network[S3]. Indeed, our HReGO exhibited a distinctly recovered *sp$^2$* network, as evident from the gradual increase in the intensity of the peak at 284.6 eV (see Figure S7)[S4,S5]. The resultant metal-like behavior of HReGO in terms of electrical conductivity (linearized I-V curve without hysteresis because of the removal of molecular defects) after the hydrazine treatment for 180 sec was clearly observed, as with the previous result (Figure S8)[S6].

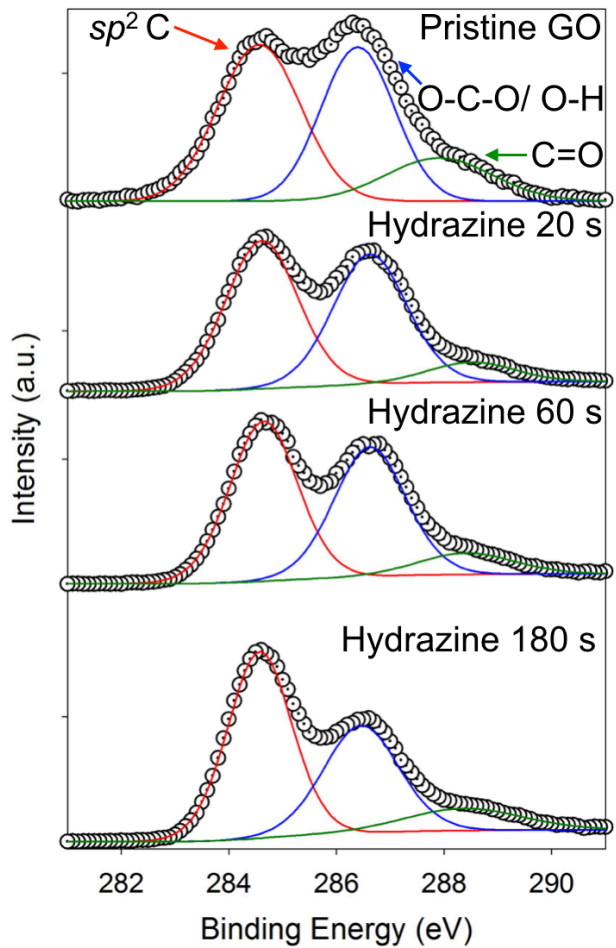

**Figure S6.** Control of broadband THz transmission (0.3 ~ 2.0 THz) by using chemical reduction of GO (i.e., hydrazine treatment). (a) XPS of GO for various hydrazine treatment times.

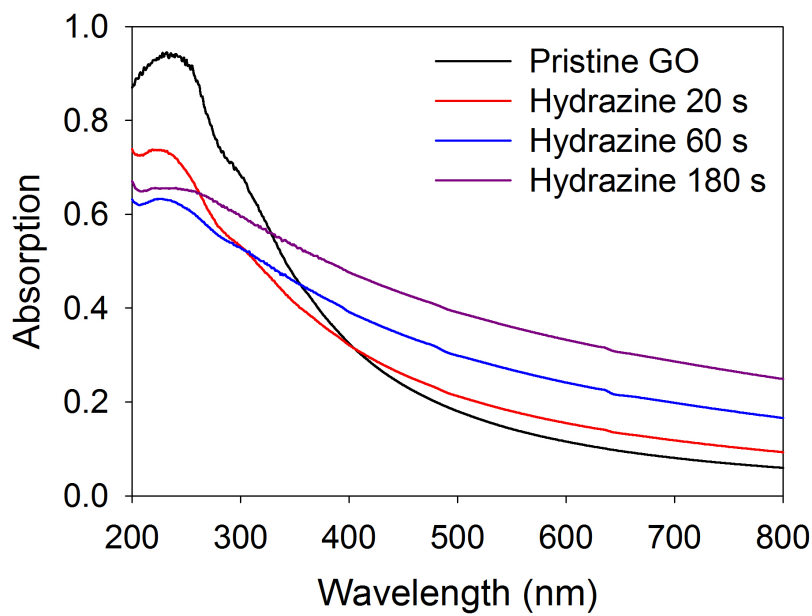

**Figure S7.** UV/VIS absorption spectra of as-prepared GO and hydrazine-treated GO. We can clearly see that the *n-pi\** shoulder peak (~ 300 nm wavelength) indicating C=O functional groups[S7] gradually disappears as the hydrazine treatment time increases.

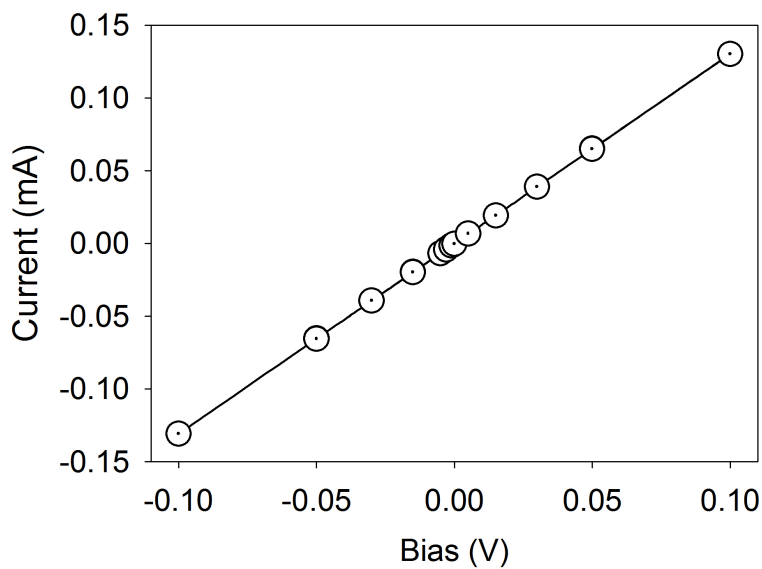

**Figure S8.** Cyclic current *versus* dc voltage measurement of hydrazine treated GO (1 min): dc Bias voltage is applied by two-terminal, interdigitated electrode with 20 μm gap. The linear profile with high current value indicates that the recovered *sp$^2$* network by hydrazine treatment transformed the insulator-like pristine GO into metal-like, reduced N-doped GO.